\documentstyle[12pt,epsf]{article}
\if@twoside
 \else
 \fi

 \parskip \medskipamount
\newcommand{\be}{\begin{eqnarray}}
\newcommand{\ee}{\end{eqnarray}}

\begin{document}

\hbox{} \nopagebreak
\vspace{-3cm} \addtolength{\baselineskip}{.8mm} \baselineskip=24pt
\begin{flushright}
{\sc OUTP-01-03P} \\
\end{flushright}

\vspace{2.5cm}
\begin{center}
{\LARGE \bf { Instanton molecules at high temperature
- the Georgi-Glashow model and beyond.}} \\
\vskip 0.3 cm 
{\large  Ian I.
Kogan$^{a,}$\footnote{kogan@thphys.ox.ac.uk},\hskip 0.2 cm 
 Alex Kovner$^{b,}$\footnote{kovner@thphys.ox.ac.uk},\hskip 0.2 cm
Bayram Tekin$^{a,}$\footnote{tekin@thphys.ox.ac.uk}
}\\
\vskip 0.3 cm  
$^a${\it Theoretical Physics, University of Oxford, 1 Keble Road, Oxford,
OX1 3NP, UK}\\ 
\vspace{0.5 cm}
$^b${\it Department of Mathematics and Statistics, 
University of Plymouth,
Plymouth PL4 8AA, UK}\\

\end{center}

\vspace*{1.5cm}


\begin{abstract}
\baselineskip=18pt  
We show that correlators of some local operators
in gauge theories  are sensitive
to the presence of the instantons even at high temperature
where the latter are bound into instanton-anti-instanton ``molecules".
We calculate correlation functions of such operators in the
deconfined phase of the 2+1 dimensional Georgi-Glashow model
and discuss analogous quantities in the chirally symmetric
phase of QCD. We clarify the mechanism by which the
instanton-anti-instanton molecules contribute to the anomaly of axial
$U(1)$ at high temperature.
\end{abstract}

\vfill

\newpage

\section{Introduction.}

In this paper we examine one particular
aspect of high temperature phase of gauge theories.
Specifically, many gauge theories at zero temperature have instantons.
The behavior of the instanton ensemble is crucial for understanding
of many dynamical properties of these theories.

In QCD with massless
fermions there are indications
that the instanton ``liquid" is responsible
for the chiral symmetry breaking \cite{dp}. A successful phenomenology
of chiral symmetry breaking has indeed been developed based on this
idea.
Another example of a gauge theory that supports instantons is the 2+1 dimensional
Georgi-Glashow (GG) model.
In this theory
(as well as in 2+1 D compact electrodynamics) the instanton-monopole
gas has been shown analytically to be responsible for confinement a
long time ago \cite{polyakov}.

When heated above a certain critical temperature these gauge
theories undergo a phase transition. In QCD the chiral symmetry is restored,
while in the GG model the deconfining transition occurs.
The status of the instantons in the hot phase
is less certain.
One thing is clear - they become less important at high temperature.
The temperature acts as an infrared cutoff on
the instanton size in QCD, suppressing the instantons of
the size greater than the inverse electric Debye screening mass \cite{gpy}.
There is also another, perhaps more significant effect.
In the GG model it has been shown analytically \cite{az, dkkt}
that the instantons at high temperatures are bound into ``molecules".
The same is believed to be true in QCD \cite{shuryak}\footnote{It
has been even
suggested that the binding of the instantons drives the chirally restoring
phase transition \cite{shuryak}. This last point
has not been definitively proven however.
Indeed, this is known not to be the
case in the GG model \cite{dkkt}, and a physically different
mechanism for the chiral transition was suggested in \cite{kkt}.
Our interest here however is not in the phase transition itself, but
rather
in the properties of the hot phase. In this respect both approaches of
\cite{shuryak} and \cite{kkt} agree that the instantons are bound in
``molecules".}.

This ``disappearance" of instantons at high temperature has been
discussed in the literature. Naively one may expect that once the
instantons are bound in pairs, their effect in the infrared is
negligible. If this is the case, the anomaly of the axial $U_A(1)$
symmetry of QCD disappears, and the chiral symmetry of QCD in the
hot phase is enhanced to $SU(N_f)\times SU(N_f)\times U_A(1)$. One
dynamical effect of such symmetry enhancement would be the
degeneracy of the correlation lengths in the scalar ($\delta$ and
$\sigma$) and the pseudo-scalar ($\pi$ and $\eta$) channels. In
fact various references to the restoration of the axial $U_A(1)$
symmetry can be found in the literature \cite{zahed}.

Of course this argumentation is too naive, and is recognized as
such by many authors in \cite{zahed}. The statement about the
anomaly of $U_A(1)$ in QCD, $\partial_\mu J^\mu_A={N_f\over
16\pi^2}\tilde F F$, is an operatorial one. As an operatorial
statement it remains true at nonzero temperature. Thus the effects
of the anomalous breaking are bound to be seen at some level even
at high temperature. Indeed the lattice data does show that the
scalar and pseudo-scalar correlation lengths are not
degenerate\cite{toussaint}. Some analytic discussion of the
consequences of the anomaly at high temperature has been given in
\cite{lee}. In particular it has been shown that in the two
flavor case the instantons contribute directly to the
correlators of the fermionic bilinears, and thus very likely lead
to splitting between the scalars and the pseudo-scalars. For
$N_f>2$ no such direct contributions exist.
 Thus the term
``$U_A(1)$ restoration" is not to be taken literally. Still since
no analytic calculation of the instanton effects at high
temperature are available, one may have lingering doubts about
their importance. It would thus be useful to have an explicit
analytic calculation of some quantities which are most sensitive
to the anomaly, and thus exhibit it in a clear way at high
temperature.

In QCD this is a difficult task due to the large number of degrees
of freedom associated with an instanton (color orientations).
However in the Georgi-Glashow model situation is much simpler,
since the instantons are not degenerate. The theory is weakly
interacting at high temperature, and thus under analytic control.
Moreover, in many respects the Georgi-Glashow model is very
similar to QCD, especially as far as the role of instantons is
concerned. The purpose of this paper is therefore to present
explicit calculations of the instanton effects at high temperature
in the Georgi-Glashow model and the close analogy between these
results and our expectations in QCD.

The general pattern of symmetries in the theories we have in mind
is the following. On the classical level the action is invariant
under a global symmetry group $G$. Quantum effects break this
symmetry anomalously down to its subgroup $H$. At zero temperature
$H$ is spontaneously broken. At high temperature $H$ is restored,
but $G$ is still broken by the anomaly. Thus a correlator $O$,
which is invariant under $H$, but not invariant under $G$, is
allowed to be non-vanishing in the anomalous theory, but has to
vanish if the anomaly of $G$ were to disappear. This particular
set of correlators is therefore likely to be very sensitive
directly to the instanton effect. We will consider examples of
such operators in the following.

 There are many parallels between QCD with massless fermions and the
 2+1 GG model. In particular
the confining physics of the latter is in many respects similar to
the chiral physics of QCD. These parallels been discussed in
detail in \cite{kkt}. The basic correspondences are the following.

$\bullet$ Classical axial $SU(N_f)\times SU(N_f)\times U_A(1)$
symmetry of QCD $\leftrightarrow$ Classical magnetic
$U_M(1)$ symmetry of the GG model .

$\bullet$ Axial anomaly due to instantons in QCD$\leftrightarrow$ Magnetic anomaly due
to monopoles in the GG model.

$\bullet$ Non-anomalous $SU(N_f)\times SU(N_f)$
subgroup
of $SU(N_f)\times SU(N_f)\times U_A(1)$  in QCD
$\leftrightarrow$ Non-anomalous $Z_{2}$ subgroup of $U_M(1)$ in the GG
model.

$\bullet$ Spontaneous breaking of  $SU(N_f)\times SU(N_f)$
by the chiral condensate
$<\bar\psi\psi>$ in QCD$\leftrightarrow$ Spontaneous breaking of $Z_2$ by
the vortex condensate $<V>$ in the GG model.

Keeping this analogy in mind, we proceed to analyze in the next
section the monopole-instanton effects in the 2+1D GG model at
high temperature. Since the theory is weakly interacting, the
calculations here are under complete analytic control. In section
3 we calculate explicitly multi-instanton contributions to relevant
correlation functions. Our results illustrate very clearly the
effects that we also expect to take place in QCD. We conclude in
section 4 with discussing further the analogy with QCD.

\section{The instantons and their interactions in the Georgi-Glashow model at high temperature.}

\subsection{The model.}
Consider the $SU(2)$ gauge theory with a scalar field in the
adjoint representation in 2+1 dimensions.
\be
S= -{1\over 2g^2}\int d^3x \mbox{tr}\left(F_{\mu \nu}F^{\mu
\nu}\right)
+ \int d^3x \left[{1\over 2}(D_\mu h^a)^2 +{\lambda
\over  4}(h^a h^a - v^2)^2 \right]
\label{model1}
\ee
In the weakly coupled
regime $v\gg g^2$, perturbatively the gauge group is
broken to $U(1)$ by the large expectation value of the Higgs field.
The photon associated with the unbroken subgroup is massless whereas
the Higgs and the other two gauge bosons $W^\pm$ are heavy with the masses
\begin{equation}
M^2_H= 2\lambda v^2, \hskip 1.5 cm M^2_W=g^2v^2.
\end{equation}
Thus perturbatively the theory behaves very much like
electrodynamics with spin one charged matter.

This theory has a global symmetry - the magnetic symmetry
\cite{thooft, kovner}.
Classically the following gauge invariant current is conserved
\begin{equation}
\tilde{F}^{\mu }=\tilde{F}^{\mu }_an^a-\frac{1}{g}\epsilon ^{\mu \nu \lambda
}\epsilon ^{abc}n_{a}({\cal D}_{\nu }n)^{b}({\cal D}_{\lambda }
n)^{c}  \label{F}
\end{equation}
with $n^a$ a unit vector in the color space in the direction of the Higgs field,
$n^a=h^a/|h|$.
This defines a conserved charge through $\Phi=\int d^2x B$. This
continuous symmetry is spontaneously broken in the vacuum, and the
massless photon is the Goldstone boson which reflects this breaking in
the spectrum.

However as is well known
the classical action eq.(\ref{model1})
supports instanton-monopole solutions with finite Euclidean action
\be
&&h^a(\vec{x})=\hat x^a h(r) \cr
&&A^a_\mu(\vec{x})= {1\over r} \left[ \epsilon_{a\mu
\nu}\hat{x}^\nu(1-\phi_1)+ \delta^{a\mu}\phi_2 +(r A-\phi_2)\hat{x}^a
\hat{x}_\mu \right]
\label{configuration1}
\ee
where $\hat x^a = x^a/r$.
In the presence of such a monopole the magnetic current is not
conserved, but rather has a non-vanishing divergence proportional to
the monopole density.
\be
\partial_\mu\tilde F_\mu={4\pi\over g}\rho
\ee
The $U_M(1)$ magnetic symmetry is thus {\it anomalous} in the quantum
theory.
As discussed in detail in \cite{kovner}
only the discrete $Z_2$ subgroup is unaffected by
anomaly and thus remains a symmetry in the full quantum theory.

The local order parameter that transforms as an eigen-operator
under the $Z_2$ magnetic symmetry is the magnetic vortex operator $V$:
\be
V\rightarrow -V
\ee
The operator $V$ has been constructed explicitly in \cite{thooft,kovner}
\begin{eqnarray}
\nonumber
V(x)&=&\exp {\frac{i}{g}}\int d^2y\ \left[\epsilon_{ij}{\frac{(x-y)_j}{(x-y)^2}
} n^a(y)E^a_i(y)+\Theta(x-y)J_0(y)\right]  \\
&=&\exp {\frac{2\pi i}{g}}\int_C dy^i
\epsilon_{ij}n^aE^a_j(y)
\label{VQCD}
\end{eqnarray}
One can think of it as
a singular $SU(2)$ gauge transformation with the field
dependent gauge function
\begin{equation}
\lambda^a(y)={\frac{1}{g}}\Theta(x-y)n^a(\vec y)
\label{lambda}
\end{equation}

The dynamical effects of the anomaly are twofold. First,
the photon becomes a pseudo-Goldstone boson and acquires
a finite albeit small mass
$m_{ph}^2\propto\exp\{-4\pi M_W/g^2\}$ (in the BPS limit) .
Another effect is confinement of $W^\pm$ bosons with the
string tension $\sigma\propto g^2 m_{ph}$.

When heated the model undergoes a phase transition to deconfined phase
at $T_c={g^2\over 4\pi}$ \cite{dkkt}.
At this transition the non-anomalous magnetic $Z_2$
symmetry is restored.

\subsection{The effective action and the instanton interactions.}

We are interested primarily in the high temperature phase, $T>T_c$.
At these temperatures the dimensional reduction applies and the effective
Lagrangian was derived in \cite{dkkt} in terms of the abelian
Polyakov line $P$.
\be
S=\int d^2x\left \{{2T\over g^2}(\partial_iP)(\partial_iP^*)
-{T\over 2g^2}M^2_D(P^2+P^{*2})+
{a^2\over 4\pi^2}\ln \left(\zeta a^2\right)(\epsilon_{ij}
\partial_iP\partial_jP^*)^2 \right \}
\label{a0}
\ee
Here the Polyakov line $P$ is a pure phase $P=\exp\{i{g\over T}A_0\}$,
and $A_0$ is the component of the vector potential parallel to the
direction of the Higgs field $n$\footnote{Note that the
transition temperature is much less than the Higgs expectation value
$T_c<<v^2$. Thus
the direction of the Higgs field is well defined
also above the transition temperature.
Of course at high enough temperature $T=O(v^2)$ the Higgs field will also
fluctuate strongly and the nonabelian part of the gauge group will
also become important. At these temperatures the theory becomes essentially
nonabelian with all the ensuing complications. To keep our calculation
under analytic control we restrict ourselves to temperatures
$T_c<T<v^2$.}.

The action eq.(\ref{a0}) deserves some explanation.
Firstly, as any dimensionally reduced Lagrangian it is
valid at distance scales greater than the inverse temperature. The
parameter $a$ appearing in it is the ultraviolet cutoff,
and is of this order $a= 2\pi/T$.
The second term expanded to order $A_0^2$ gives the electric
Debye screening mass to $A_0$. The mass squared is proportional to the
fugacity of the charged particles, which in this model are the $W^\pm$
bosons
$M^2_D\propto\exp\{-M_W/T\}$. The exact proportionality coefficient is
straightforwardly calculable at one loop,
but is inessential for our purposes.
The last term is induced by the monopole-instantons.
As discussed in \cite{dkkt} the monopoles in this effective action
appear as vortices of the Polyakov line $P$ with unit winding.
Indeed the last term in eq.(\ref{a0}) is of the type of Skyrme term
and vanishes on any smooth configuration of $A_0$.
Its only contribution comes from points at which the winding of $P$
is concentrated.
It therefore just counts the number of vortices and anti-vortices of $P$,
that is of monopoles and anti-monopoles. On a configuration with one
monopole the contribution of this term to the
partition function
is equal to $\zeta$. Thus $\zeta$ is the fugacity of the
monopole (or anti-monopole), $\zeta\propto \exp\{-4\pi M_w/g^2\}$
\footnote{The proportionality constant is slightly different
than at zero temperature
reflecting the fact that the field modes with momenta $p>T$ have been
integrated out to arrive at the effective
action eq.(\ref{a0})\cite{az}.}.
An important point is that this dimensionally reduced action
has not been derived in the derivative expansion, but rather in
the expansion
in powers of the exponentially small parameters $M_Da$
and $\zeta$ \cite{dkkt}. It is thus valid
for all distance scales above the UV cutoff $a$.

It is easy to see from the effective action eq.(\ref{a0}) that the
monopoles are bound by a (screened) linear potential.
Consider a configuration
with unit winding of the field $P$. Due to the potential term $P^2$
the minimal action configuration can not be a rotationally symmetric
hedgehog. Such a configuration would ``cost" action proportional
to the volume, since the field $P$ would be away from its vacuum value
everywhere in space. The best one can do is to have a quasi one dimensional
strip in which the winding is concentrated, while everywhere else in space
$P$ would be equal to $1$ (or $-1$). This configuration is schematically
depicted on Fig. 1.
\begin{figure}
\begin{center}
\epsfxsize=2.5in
\epsfbox{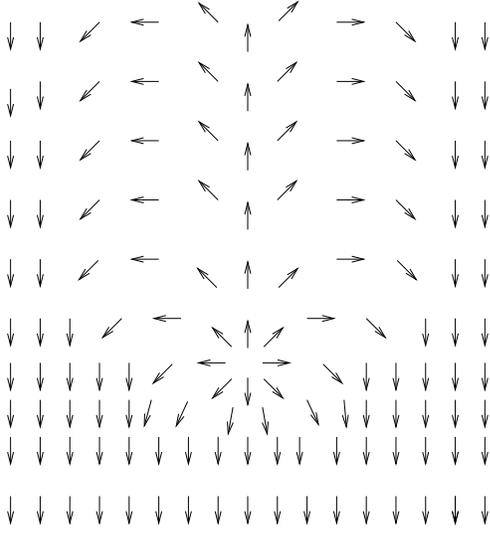}
\caption{The  string-like configuration of the Polyakov line $P$
that accompanies the monopole-instanton in the high temperature
phase.}
\end{center}
\end{figure}
The width of the ``dual confining string" must be clearly of
order of $M^{-1}_D$, while the action per unit length
$s\propto {T\over g^2}M_D$. Thus the action of a single monopole
diverges linearly with the size of the system. Obviously the
monopole-anti-monopole pair separated by a distance $L$
have the action $Ls$. When the distance is large
enough, another pair can be popped from the thermal ensemble
and screen the linear potential. The critical distance at which
this happens
is clearly determined by\footnote{In this equation we neglected the
effects of the finite temperature ``renormalization" of the monopole
fugacity $\zeta$.}
\be
L_cs=2\log\zeta={8\pi M_W\over g^2}
\ee
or
\be
L_c\propto {M_W\over T}M^{-1}_D
\ee
Thus as long as the temperature is much lower than $M_W$, the length
of the ``dual string" is much greater than its thickness.
The ``potential" between the monopoles is therefore linear, but screened
at large distances, much in the same way as the confining potential in gauge
theories with heavy fundamental charges\footnote{Of course we should
always keep in mind that while talking about instanton ``potential''
we really mean ``action'' and not energy. Thus physically
``confinement'' of instantons is very different that confinement of
charged particles.}.

Now it is interesting to consider the correlation functions of the
vortex operators. There is a direct analogy between these
correlation functions and the correlators of the fermionic
bilinears in QCD. In particular if the  magnetic $U_M(1)$
were restored at high temperature, this would imply a simple
relation between correlation lengths in different channels, 
directly analogous to the would be
degeneracy between the $\eta$ and the pions in QCD.
Recall that the vortex operator in the GG model is a
pseudo-scalar\cite{kovner},
 so that the
parity transformation acts on it as
\be
V\rightarrow V^*
\ee
The parity even and odd eigenstates are therefore
\be
V^+={V+V^*\over 2},\ \ \ \ \ \ \  V^-=i{V-V^*\over 2}
\label{parity}
\ee
If the magnetic $U_M(1)$ is restored, the correlators $<VV>$ and $<V^*V^*>$
both vanish, and thus the following relation should hold
\be
<V^+(x)V^+(y)>=<V^-(x)V^-(y)>
\ee
The correlation lengths in the parity even and the parity odd
channels should be the same. This is the direct analog of
the would be
equality of the correlation lengths in the $\eta=\bar\psi\gamma_5\psi$ and
the $\sigma=\bar\psi\psi$ channels in QCD.
We will see in the following that this equality in fact does not hold due to the
instanton effects.

\section{Instanton effects at high temperature.}

\subsection{$U_M(1)$ invariant correlators of $V$.}

Let us remind the reader how calculation of the correlators of $V$
proceeds in the framework of the effective action eq.(\ref{a0}).
For pure gauge theories this has been discussed in detail in \cite{kaks}.
In the present, essentially Abelian case the procedure is very similar.
The vortex operator $V(x)$ in eq.(\ref{VQCD}) induces a singular
abelian gauge
transformation which forces the vector potential $A_0$ to jump by $\pi$
across the curve $C$ in the definition eq.(\ref{VQCD}). Thus the Polyakov
line $P$ is forced to change sign across $C$. Formally,
the insertion of the operator $V(x)$ into the Euclidean
finite temperature path integral leads to the shift
\be
\partial_iA_0\rightarrow\partial_iA_0-a_i
\label{shift}
\ee
in the derivative terms in
eq.(\ref{a0}). Here $a_i$ is a ``vector potential" of
a point-like vortex of vorticity $1/2$ at the point $x$
\be
a_i(y)=\pi n^C_j(y)\delta(y-C)
\label{a}
\ee
where $n^C(y)$ is a unit vector normal to the curve $C$ at the point $y$.
The result does not depend on the curve $C$, but only on its end point $x$,
since the shape of the curve may be changed into $C'$ by
transforming $P\rightarrow-P$ in the area enclosed by $C-C'$\cite{kaks}.

The calculation of the correlation function $V(x)V^*(y)$
leads to the path integral with the vortex at $x$ and the
anti-vortex at $y$.  At high enough temperature
due to the factor $T/g^2$ in the action this path integral
can be calculated in the steepest descent approximation.
The integral then is dominated by the solution of the classical
equations of motion subject to the condition that $P$ has to change sign
across the curve $C$ that connects the points $x$ and $y$, see Fig.2.

\begin{figure}
\begin{center}
\epsfxsize=3.0in
\epsfbox{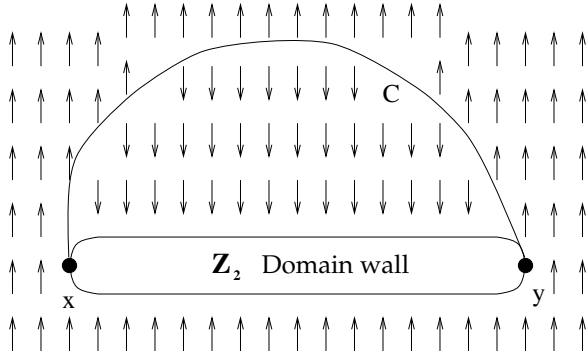}
\caption{The configuration of the Polyakov line $P$
that dominates the correlator $V^*(x)V(y)$.}
\end{center}
\end{figure}
The shape of this solution is easy to understand \cite{kaks}. In the
hot phase the configurations $P=1$ and $P=-1$ are degenerate. There
are therefore classical solutions which have the form of domain walls,
that interpolate between these two configurations, the so called $Z_2$
domain walls \cite{chris}.
Since our effective action is a simple sine-Gordon theory, the solution
for such a domain wall is known explicitly\footnote{The Skyrme term
in eq.(\ref{a0}) does not play any role in this discussion, since the
solution in question is smooth and this term vanishes
for smooth functions $A_0$.}
\be
A_0= {2\pi T\over g}\mbox{arctan}\,e^{M_D(x-x_0)}
\ee
The action of this solution per unit length is
\be
\tilde\sigma={8T\over g^2}M_D
\label{s}
\ee

In the calculation of $V(x)V^*(y)$
the discontinuity in the boundary conditions forces $P$
to jump from the vicinity of one ``vacuum" to the vicinity of the other
across $C$.
But since $P$ has to go to the same value in all directions at infinity,
it has to return to the original vacuum.
The solution thus is a domain wall of Fig.2.
Using the result eq.(\ref{s})
we find that for large separations\footnote{Again the Skyrme term does not
contribute, since the contribution of the cut $C$ is subtracted explicitly
by the shift eq.(\ref{shift}).}
\be
<V(x)V^*(y)>\propto\exp\{-|x-y|\tilde\sigma\}
\ee
Actually one can do a little better and determine the prefactor
quite easily. The prefactor originates from the action
associated with the endpoints $x$ and $y$ at which the wall terminates.
Close to the endpoints (at distance smaller than $M_D^{-1}$) the mass
term is irrelevant and the configuration
of $A_0$ must be that of a rotationally symmetric vortex (anti-vortex)
of vorticity $\pi$
\be
A_0(z)={\theta(z)\over 2}
\ee
with $\theta$ a planar angle relative to the location of the vortex.
Such a vortex carries the Coulomb ``self energy"
\be
\delta S={T\over 2g^2}\pi\int_{a}^{M^{-1}_D}{dr \over r}=-{\pi T\over 2g^2}
\ln (aM_D)
\label{self}
\ee
Adding the contribution of both end points we find\footnote{In principle
we should also subtract $2M_D^{-1}$
from the length of the string for consistency. This correction is however
sub-leading,
since it brings in a factor  $\exp\{2\tilde\sigma M_D^{-1}\}$ which is
of order 1
in the interesting temperature range.}
\be
<V(x)V^*(y)>=(aM_D)^{\pi T\over g^2}\exp\{-|x-y|\tilde\sigma\}
\label{corr}
\ee

Note that this calculation was performed entirely in the zero
instanton sector.
There is of course an instanton-anti-instanton contribution. We will discuss it
at the end of this section.

\subsection{Instanton sensitive correlators.}
Let us now consider the calculation of $<V^2(0)>$, which is
the simplest operator not invariant
under the magnetic $U_M(1)$, but still invariant under the non-anomalous
$Z_2$.
Neglecting the instanton contributions, we should find
that the VEV of this operator vanishes.
Indeed the path integral for $<V^2(0)>$
has the boundary condition imposing vorticity $2\pi$ of $P$ at the
point $x=0$. In the zero instanton sector the action with such
boundary condition diverges. The reason is precisely the same as the
one for linear potential between the instantons. The external
vortex associated with $V^2(0)$ will pick up a ``dual confining string",
which carries finite action per unit length. This configuration
has action proportional to the linear size of the system and thus
the VEV of $V^2$ vanishes exponentially in the thermodynamic limit.
The leading non-vanishing contribution to $<V^2(0)>$ comes from the
sector with one anti-instanton .
Clearly if there is an anti-instanton in the vicinity of the point $x=0$,
the dual string that originates at $x=0$ will terminate on it
and thus the result will be finite.
To calculate the expectation value we note that
it is dominated by the anti-instanton sitting relatively close to
the point $x=0$,
since otherwise the action of the configuration is very large.
On the other hand if the coordinate of the anti-instanton $x_a$
is smaller than $M_D^{-1}$, the
action of the classical solution is just the
action of a dipole with the vector potential
\be
a_0(x)=\theta(x)-\theta(x-x_a)
\ee
The expectation value is therefore given by the integral over the
quasi zero mode $x_a$
\begin{eqnarray}
<V^2>&=&\zeta\int_{a<|x_a|<M^{-1}_D} {d^2 x_a\over \pi a^2}\exp\{-S(a_0)\}
\nonumber \\
&=&
\zeta a^{-2}\int d(l^2)\exp\{-{4\pi T\over g^2}
\ln \left({l^2\over a^2}\right)\nonumber\\
&=&
\zeta{g^2\over 4\pi T-g^2}[1-(M_Da)^{2{4\pi T-g^2\over g^2}}]
\end{eqnarray}

Let us now consider the correlation function $<V(x)V(y)>$. As far
as the magnetic quantum numbers are concerned, this correlator is
the same as the composite operator $V^2$, but it is more
interesting since it must also exhibit some physical correlation
length. The calculation of this quantity is not much different. In
the zero instanton sector it vanishes for the same reason as
$V^2$. It carries a net ``vorticity" and this vorticity has to be
screened by an anti-instanton in order to give a finite result. In
principle the anti-instanton can be located at an arbitrary point
in space. The field configuration corresponding to two external
vortices associated with insertions of $V$ and the anti-instanton
is schematically depicted on Fig.3.

\begin{figure}
\begin{center}
\epsfxsize=4in
\epsfbox{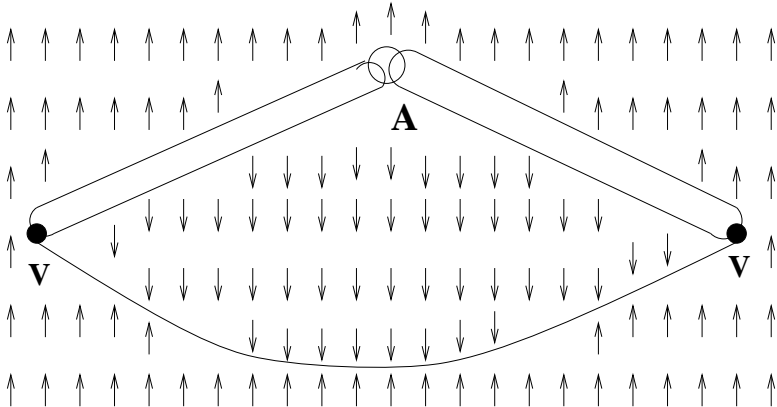}
\caption{The configuration of one anti-instanton contributing to
the correlator $V(x)V(y)$.}
\end{center}
\end{figure}

Clearly the contribution of
anti-instantons sitting at a significant distance from the straight
line connecting $x$ and $y$ is exponentially suppressed, since the
``dual confining string" in this configuration is longer. Thus the
leading contribution is given by the anti-instantons sitting within
the transverse distance $M_D^{-1}$ of the straight line. The
correlation function thus is \be <V(x)V(y)>=\zeta\int {d^2z\over
\pi a^2} \exp\{-S(x,y,z)\} =\zeta\int_x^y {dz_1\over \pi M_Da^2}
\exp\{-S(x,y,z_1)\} \ee where we have taken  $x=(x_1,0)$ and
$y=(y_1,0)$. Here $z$ is the position of the anti-instanton. As
long as $|x-y|>>M_D^{-1}$, the main contribution comes from the
anti-instantons far from the endpoints $x$ and $y$. For these
configurations the integral over $z_1$ just gives $|x-y|$, and the
action is the sum of the action of the string $\tilde\sigma|x-y|$
plus the self energies of the anti-instanton and the vortices at the
endpoints $x$ and $y$. Those are calculated as in
eqs.(\ref{self},\ref{corr}). So that finally we get \be
<V(x)V(y)>=\zeta(aM_D)^{{3\pi T\over g^2}-1}{|x-y|\over a}
\exp\{-|x-y|\tilde\sigma\} \label{corr1} \ee Interestingly enough
the correlation length that appears in this correlation function
is the same as the one appearing in the correlator of vortex and
anti-vortex $V(x)V^*(y)$.

Thus as expected, both the VEV of $V^2$ and the correlator $V(x)V(y)$
are sensitive to the instantons. Both these quantities are
indeed directly proportional to the fugacity of the instanton $\zeta$.
Even though in the $U_M(1)$ invariant sector the instantons appear only
in bound pairs, in the calculation of the $U_M(1)$ non-invariant
correlation functions they do appear separately. Such a single
instanton (anti-instanton)
binds to the external insertion that carries
the explicit $U_M(1)$ charge and screens this charge. This
leads to non-vanishing correlation functions in the channels whose
quantum numbers match those of an integer number of instantons
(anti-instantons).

\subsection{Multi-instanton contributions.}

As mentioned earlier, there are also multi-instanton contributions to the
vortex correlation functions. Those become important
when the separation $|x-y|$ is large enough. The reason is that although
each extra instanton-anti-instanton pair is suppressed by $\zeta^2$,
at large $|x-y|$ the ``entropy" is large enough and thus these contributions
are not negligible.

Consider first the correlator $V(x)V^*(y)$. In order that the multi-instanton
contribution not be suppressed exponentially, all instantons must
be located on the straight line connecting the points $x$ and $y$. Moreover,
the instantons and anti-instantons must alternate, since otherwise
some sections of the dual confining string  connect regions of
vorticity 3, and thus has much higher dual string tension.
Other than these constraints, the coordinates of instantons and anti-instantons
are arbitrary and have to be integrated over.
The contribution of $n$ instantons and $n$ anti-instantons 
is therefore given by
\be
<V(x)V^*(y)>_n&=&(aM_D)^{\pi nT\over g^2}\exp\{-|x-y|\tilde\sigma\}\\
&\times& \zeta^{2n}(aM_D)^{4\pi nT\over g^2}{ \pi
M_Da^2}^{-2n}\int_x^ydy_n\int_x^{y_n}dx_n\int_x^{x_n}dy_{n-1}...
\int_x^{y_1}dx_1\nonumber \ee Here $x_i$ and $y_i$ are
longitudinal coordinates of instantons and anti-instantons
correspondingly \footnote{Note that the usual symmetry factor
${1\over n!}{1\over n!}$ does not appear in this expression due to
ordering of the coordinates of the instantons and
anti-instantons.}. Performing the integration and summing over $n$
we find
\be <V(x)V^*(y)>=(aM_D)^{\pi T\over
g^2}\exp\{-|x-y|\tilde\sigma\} \cosh\{\zeta(aM_D)^{{2\pi T\over
g^2}-1}{|x-y|\over \pi a}\} \label{corr2} \ee

The multi-instantons have therefore a very nontrivial effect on the vortex
correlator. They ``split" the correlation length in two, generating two
distinct correlation lengths
\be
\tilde\sigma_\pm=
\tilde\sigma\pm\zeta(aM_D)^{{2\pi T\over g^2}-1}{1\over \pi a}
\ee

The calculation of the multi-instanton contributions to $V(x)V(y)$
is very similar, and gives
\be
<V(x)V(y)>=(aM_D)^{\pi T\over g^2}\exp\{-|x-y|\tilde\sigma\}
\sinh\{\zeta(aM_D)^{{2\pi nT\over g^2}-1}{|x-y|\over \pi a}\}
\label{corr3}
\ee
Again we see that the same set of correlation lengths appears
in the two correlation functions.

Eqs.({\ref{corr2},\ref{corr3}) are very simple from the point of view
of the parity eigenstates $V^+$ and $V^-$ eq.(\ref{parity}). They can
be rewritten as
\be
<V^+(x)V^+(y)>&=&(aM_D)^{\pi T\over g^2}\exp\{-|x-y|\tilde\sigma_-\}\nonumber \\
<V^-(x)V^-(y)>&=&(aM_D)^{\pi T\over g^2}\exp\{-|x-y|\tilde\sigma_+\}
\ee
Thus we see explicitly that the correlation lengths in the scalar and
pseudo-scalar
channels are not equal due to the instanton effects.
This difference decreases with temperature, but in the temperature range 
under consideration ($T\ll M_W$) always remains finite. Recalling that 
$M_D^2\propto\exp\{-M_W/T\}$
and $\zeta\propto \exp\{-4\pi M_W/g^2\}$ we find that the ratio 
of the difference of the inverse correlation lengths to their sum varies
between
\be
{\tilde\sigma_+-\tilde\sigma_-\over\tilde\sigma_++\tilde\sigma_-}=
\exp \{-{\pi M_W\over g^2}\}, \ \ \ \ \ T\rightarrow T_c={g^2\over 4\pi}
\ee
and
\be
{\tilde\sigma_+-\tilde\sigma_-\over\tilde\sigma_++\tilde\sigma_-}=
\exp \{-{5\pi M_W\over g^2}\}, \ \ \ \ \ T\rightarrow M_W
\ee

Finally it is worth mentioning that the quantities that get the
direct instantons contributions naturally
are proportional to the instanton fugacity, but their smallness
is not uniform.
For example as already mentioned, the correlation lengths in the vortex-vortex
channel is the same as in the vortex-anti-vortex channel.
Moreover comparing the
vortex - vortex correlator eq.(\ref{corr3}) with the vortex-anti-vortex correlator
eq.(\ref{corr2}) we see that although the former is smaller at short
distances , at distances of order
$\pi a\zeta^{-1}(aM_D)^{1-{2\pi T\over g^2}}$
the two are numerically practically equal.

\subsection{$N>2$.}
Let us briefly comment on the situation in the GG type models for
$N>2$. The quantitative discussion of the phase transition will be
given elsewhere \cite{inprep}. Here we want to discuss only the
main qualitative features. The basic degree of freedom is still a
vortex operator $V(x)$, and the effective theory at high
temperature is similar to eq.(\ref{a0}) with the main difference
that the potential term is $P^N+P^{*N}$, and thus is $Z_N$
invariant\footnote{The actual situation generically is more
complicated than that. In particular there are $N-1$ distinct
vortex operators. Also there are $N-1$ distinct Abelian Polyakov
lines, corresponding to the $N-1$ Cartan algebra generators, and
all of those should appear in the effective action. However one
can always choose the parameters of the model to be such that one
of these operators is much lighter than the others. The rest of
the Polyakov lines then can be integrated out and the effective
action depends only on one $P$. For simplicity we consider only
this situation  in the present article, although the main
conclusions are the same also in the generic case.}. The instanton
again is just a vortex of $P$ with unit vorticity.

First consider the calculation of $<VV^*>$. Just like for $N=2$,
the contribution in zero instanton sector is given by the $Z_N$
domain wall
 \be
 <V(x)V^*(y)>\propto\exp\{-|x-y|\tilde\sigma\}
 \ee
 The instanton-anti-instanton contribution now is however
 different. The point is that the instanton (anti-instanton), having winding number
 one, is the source (sink) of $N$ $Z_N$ domain walls. 
Thus the system of the vortex
 $V(x)$ plus an anti-instanton at some point $x_1$ together are the
 sink of $N-1$ domain walls. The string tension of this bunch is
 roughly $(N-1)\tilde \sigma$. Thus in order for the instanton-anti-instanton
 contribution not to be suppressed by the exponential factor, the anti-instanton must sit
 within a distance $[(N-2)\tilde\sigma]^{-1}$ of the instanton, see
Fig. 4.

\begin{figure}
\begin{center}
\epsfxsize=4.5in
\epsfbox{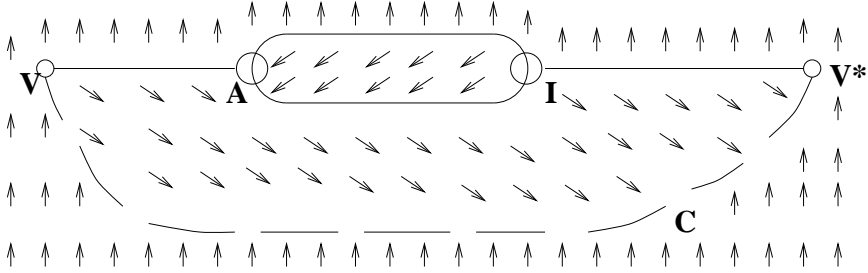}
\caption{The instanton-anti-instanton configuration contributing to
the correlator $V(x)V^*(y)$ for $N=3$. The instanton (anti-instanton)
is a source (sink) of three elementary $Z_3$ domain walls. The curve
$C$ is the ``fake'' infinitely thin wall as in eq.(\ref{a}).}
\end{center}
\end{figure}

Again summing over the number of the AI pairs, we find that the
contribution exponentiates

\be <V(x)V^*(y)>=(aM_D)^{4\pi T\over
g^2N^2}\exp\{-|x-y|\tilde\sigma\} \exp\{\zeta^2(aM_D)^{{2\pi
T\over g^2}-1}{|x-y|\over \pi a^2(N-2)\tilde\sigma}\} \ee

So the result is quite distinct from the $N=2$ case. Only one
correlation length is present, although there is indeed a direct
contribution of instantons into this length of order $O(\zeta^2)$.

The behavior of the correlator $<V(x)V(y)>$
 is also very different. Since this correlation function is not
 symmetric under the magnetic $Z_N$ symmetry, which is unbroken in
 the high temperature phase, it strictly vanishes. Instead, the
 instantons contribute to the correlation functions with the
 quantum numbers of $V^N$. In particular consider
 $<V(x)V^{N-1}(y)>$. The leading contribution to this correlation
 function comes from the configuration with an anti-instanton within
 the distance  $[(N-2)\tilde\sigma]^{-1}$ of the point $y$ and an
 arbitrary number of the AI pairs.
 \be
<V(x)V^{N-1}(y)>= \zeta(aM_D)^{{(4+N^2)\pi T\over N^2g^2}-1}
\exp\{-|x-y|\tilde\sigma\} \exp\{\zeta^2(aM_D)^{{2\pi T\over
g^2}-1}{|x-y|\over \pi a^2(N-2)\tilde\sigma}\}
\ee

In this calculation we only kept the leading contribution to the
correlation function, which in the absence of instantons decays
with the correlation length $l^{-1}=\tilde\sigma$. There is of
course another contribution. The configurations where the
instantons and anti-instantons are allowed to separate will lead to
the component with the correlation length
$l^{-1}=(N-1)\tilde\sigma$.  This is precisely the correlation
length that in the absence of instantons dominates the correlator
$<V^{N-1}(x)V^{*(N-1)}(y)>$. The instantons clearly lead to the
mixing of the operators $V$ and $V^{*(N-1)}$ in full analogy with
the case $N=2$. This mixing is accompanied with the instanton
induced increase in the longest correlation length

\be
[\tilde\sigma]^{-1}\rightarrow[\tilde\sigma- (aM_D)^{{2\pi
T\over g^2}-1}{\zeta^2\over \pi a^2(N-2)\tilde\sigma}]^{-1}
 \ee

Thus for $N>2$ the instantons do not remove the degeneracy between
$V+V^*$ and $V-V^*$,but they do significantly affect other
correlation functions consistently with the magnetic $Z_N$
symmetry.

We conclude that even though at high temperature the instantons
are bound in pairs, they are still relevant in the infrared. The
anomaly does not disappear. Its manifestation is non-vanishing of
non-singlet correlators as well as the difference of the correlation
lengths in the scalar and pseudo-scalar channels for $N=2$ and the
operator mixing consistent with the unbroken magnetic $Z_N$ for
$N>2$.

\section{Discussion}
The calculation presented in this paper illustrates two simple
points. First, which correlation functions vanish and which don't
is determined entirely by the non-anomalous symmetries. Second, the
instanton contributions are important at high temperature even
though the instantons in the partition function are bound in
pairs. Both these features we expect of course to hold also in
QCD.

In particular let us consider the fermionic bilinear correlation
functions. The fermionic bilinears that transform as irreducible
representations of the non-anomalous $SU(N_f)\times SU(N_f)$ are
\be O^{ij}(x)=\bar\psi^i_L(x)\psi^j_R(x), \ \ \ \ \ O^{\dagger
ij}(x)=\bar\psi^i_R(x)\psi^j_L(x)\ee
whereas the parity eigen-operators are
\be O_{\pm}=O\pm O^{\dagger}\ee

Now consider the correlation functions $O_+(x)O_+(y)$ and
$O_-(x)O_-(y)$. If the correlators $T=<O(x)O(y)>$
 and $<T^\dagger=O^\dagger(x)O^\dagger(y)>$ both vanish, then the
 correlation functions of the parity even and the parity odd
 operators are equal. Whether this is true or not depends on the
 number of flavors. For $N>2$ the vanishing of these correlation
 functions is assured by the $SU(N_f)\times SU(N_f)$ symmetry,
 since $T$ is the product of two fundamental representations
 (separately for the right and left symmetries) and therefore does
 not contain a singlet. For $N_f=2$ on the other hand the product
 of two fundamentals does contain a singlet. Thus there is no
 symmetry obstruction for $T$ to be nonzero\footnote{In this discussion
 we did not mention at all the anomalous
$U_A(1)$. Of course, if we assume that it is restored, the parity
odd and even channels must be degenerate, since they transform
into each other under the axial $U(1)$.}.

Dynamically the contributions to $T$ come from the instanton
sector. For $N_f=2$ the number of fermionic operators in $T$ is
just right to saturate the fermionic zero modes of an instanton.
Recall that the fermionic zero modes at high temperature away from
the core of the instanton exponentially decrease  as $\exp\{-\pi T
|x|\}$ \cite{gpy,dp}. This is the same exponential decay as of the
lowest fermionic Matsubara mode. Thus we expect the situation to
be very similar to the $N=2$ Georgi-Glashow model. The
perturbative behavior of the correlator $O(x)O^\dagger(y)$ is

\be <O(x)O^\dagger(y)>\propto \exp\{-2\pi T |x-y|\} \ee

The instanton-anti-instanton contributions have the same
exponential behavior, and extra powers of $|x-y|$ due to the
arbitrariness of the instanton's (anti-instanton's) position. It is
very likely that the sum over the number of instantons
exponentiates as in eq. (\ref{corr2}) and results in the splitting
of the basic correlation length into two $2\pi T\pm x\zeta$. These
two distinct correlation length then appear separately in the
parity even and parity odd channels.

For $N_f>2$ the number of fermions in $T$ does not match the
number of the fermionic zero modes in the instantons, and thus the
only way instantons can contribute is through AI pairs. They will
however give direct contributions to the correlators of the type
$T_N=<O(x)O_{2N-2}(y)>$ where $O_{2N-2}$ is the operator
containing $2N-2$ fermions, such that the quantum numbers of $T_N$
are the same as of 't Hooft's determinant. This is essentially the
argumentation of \cite{lee}.  Again this is mimicked perfectly by
our discussion of the $N>2$ GG model.

Although analytic multi-instanton calculations in QCD are
difficult, the toy GG model gives us a very clear picture of what
the expected result is. Hopefully this knowledge may give us a
hint of how to separate the most important contributions so that
an analytic calculation may become possible after all.

\leftline{\bf Acknowledgments}
 A.K. is supported by PPARC. The research of  I.K. and B. T. are
supported by   PPARC Grant PPA/G/O/1998/00567.

\end{document}